\documentclass[]{spie}  

\usepackage{amsmath,amsfonts,amssymb}
\usepackage{graphicx}
\usepackage[colorlinks=true, allcolors=blue]{hyperref}
\usepackage[table]{xcolor}
\usepackage{xspace}

\DeclareRobustCommand{\ion}[2]{%
\relax\ifmmode
\ifx\testbx\f@series
{\mathbf{#1\,\mathsc{#2}}}\else
{\mathrm{#1\,\mathsc{#2}}}\fi
\else\textup{#1\,{\mdseries\textsc{#2}}}%
\fi}

\def\utw{\smash{\rlap{\lower5pt\hbox{$\sim$}}}}
\def\udtw{\smash{\rlap{\lower6pt\hbox{$\approx$}}}}

\title{Arcus X-ray telescope performance and alignment}

\author[a]{Hans Moritz G\"unther}
\author[b]{Peter Cheimets}
\author[a]{Eric D. Miller}
\author[c]{Casey DeRoo}
\author[b]{Randall K. Smith}
\author[d]{Andrew Ptak}
\author[a,e]{Ralf K. Heilmann}
\affil[a]{MIT Kavli Institute for Astrophysics and Space Research, Cambridge, MA 02139, USA}
\affil[b]{Center for Astrophysics, Harvard-Smithsonian Astrophysical Observatory, Cambridge, MA 02138, USA}
\affil[c]{Dept. of Physics \& Astronomy, University of Iowa, Iowa City, IA 52242, USA}
\affil[d]{NASA Goddard Space Flight Center, Greenbelt, MD 20771, USA}
\affil[e]{Space Nanotechnology Laboratory, MIT Kavli Institute for Astrophysics and Space Research, Cambridge, MA 02139, USA}

\authorinfo{Send correspondence to H.M.G. (E-mail: hgunther@mit.edu)}

\begin{document}
\maketitle

\begin{abstract}
Arcus is a concept for a probe class mission to deliver high-resolution FUV and X-ray spectroscopy. For X-rays, it combines cost-effective silicon pore optics (SPO) with high-throughput critical-angle transmission (CAT) gratings to achieve $R> 3000$ in a bandpass from 12-50 \AA. We show in detail how the X-ray and the UV spectrographs (XRS and UVS) on Arcus will be aligned to each other. For XRS we present ray-tracing studies to derive performance characteristics such as the spectral resolving power and effective area, study the effect of misalignments on the performance, and conclude that most tolerances can be achieved with mechanical means alone.
We also present an estimate of the expected on-orbit background.

\end{abstract}


\section{INTRODUCTION}
\label{sect:introduction}
High-resolution X-ray and UV spectroscopy open a window into the physics of the universe that often cannot be observed with any other technique. X-rays and UV can probe the hottest and most ionized gas that remains invisible in longer wavelengths because the high ionization levels do not produce observable transitions in the radio, infrared or optical. Thus, observations in the UV and X-rays probe the most energetic processes in a number of systems. A few examples are: stellar space weather \cite{2022AN....34320019B}, the accretion onto young stars, where the UV and X-rays come from the energetic infall of material from the disk onto the stellar surface; the innermost regions of the accretion disks around black holes; and the absorption of background (X-ray) light from distant, bright continuum sources by the warm-hot inter-galactic medium to find the missing baryons\cite{10.1117/12.2231193} - see Smith et al.\ and France et al.\ in this volume for more details. High-resolution UV and X-ray spectroscopy that resolves the profiles of individual emission or absorption lines is particularly valuable, because it allows us to address a host of physical questions that cannot be answered by simply measuring the broad-band X-ray flux.
To follow-up with one specific example: in the case of accretion onto young stars, resolving the kinematic line profile of emission in different ions can tell us which part of the emission is formed in the infalling, red-shifted accretion column, which part is related to the stellar corona seen at the rest wavelength of the star (possibly with small velocities from up and downflows in the corona), and which part is formed in (or possibly absorbed by) the blue-shifted outflow.

The Arcus mission is a concept that will address those challenges with two instruments for X-ray and UV high-resolution spectroscopy. The mission evolved through several stages. It was originally
proposed as an instrument mounted on the International Space Station\cite{10.1117/12.2062671} and then redesigned as a satellite\cite{10.1117/12.2231778,10.1117/12.2272818}. Arcus' X-ray spectrograph (XRS) will perform
high-resolution spectroscopy in the soft X-ray range (about 12-50~\AA{}) with a resolving power $R>3000$ and an effective area $A_\textrm{eff}>
300\;\mathrm{cm}^2$ for most of the bandpass with a peak close to $A_\textrm{eff} > 600\;\mathrm{cm}^2$ around 18~\AA{}. The resolving power is a factor of
$3-5$ larger than what existing instruments on Chandra or XMM-Newton can
deliver and the effective area is also significantly higher; the exact number
depends on the bandpass, e.g.\ in the crucial region around the O~{\sc vii}
He-like triplet, which is density and temperature sensitive, Arcus will reach
about two orders of magnitude more effective area than Chandra/HETG currently
has. The UV spectrograph (UVS) is described by Fleming et al.\ in this volume and the UV science case is discussed by France et al.\ in this volume. In this work, we concentrate on the XRT.


\section{CO-ALIGNMENT OF THE X-RAY AND UV INSTRUMENTS}
\label{sect:alignment}
\begin{figure}
    \begin{center}
      \includegraphics[width=0.3\textwidth]{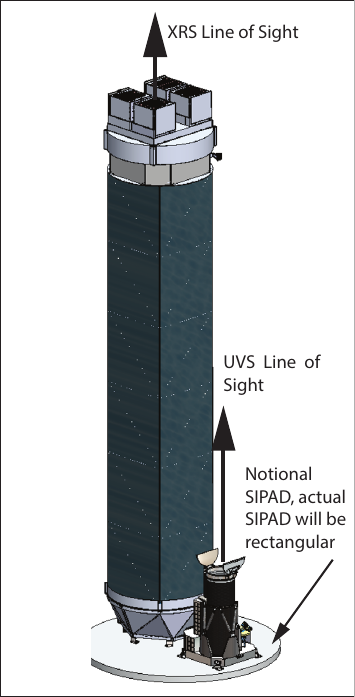}
    \end{center}
    \caption{XRS and UVS mounted on the SIPAD. Each instrument LOS is shown.\label{fig:1}}
\end{figure}

Arcus Probe is a mission that consists of 2 separate spectrometers (see Figure~\ref{fig:1}) operating in concert to acquire simultaneous spectra at X-ray and UV wavelengths. To succeed, the spacecraft pointing stability must meet the requirements of the spectrometer with the tighter pointing requirements, and both spectrometers must have a given target in their fields of view (FOVs) at the same time.

The two spectrometers are:
\begin{description}
    \item[The X-ray Spectrometer (XRS):] This is a photon counting soft X-ray spectrometer that operates in the 12Å - 50Å spectrum using silicon pore optics mirror modules (SPO MM, or just SPOs) and Critical-Angle Transmission (CAT) gratings to collect (SPOs), focus (SPOs), and diffract (CATs) photons onto a detector array 12m away.
    \item[The UV Spectrograph (UVS):] This is a photon counting far UV spectrometer that operates in the 1020Å-1540Å spectrum using an off-axis Cassegrain telescope and a curved reflection grating to collect (Cassegrain telescope), focus (Cassegrain telescope), and diffract (curved grating) photons onto a photon counting detector.
\end{description}
The spectral images that both instruments create are reconstructed on the ground after the observation is complete. This is done by augmenting the photon arrival information with the Observatory pointing and instrument aspect information to remove the effects of pointing error and instrument distortion. UVS’s spectral resolution requirement dictates a tighter pointing requirement than what XRS requires. Thus, the Observatory is configured to point UVS, and the XRS is designed to be internally aligned to the UVS.

\subsection{Arcus Probe Alignment, an overview}
It is critical to ensure that the UVS and XRS are pointed at the same objects throughout the science operations of the Probe mission. The UVS pointing direction, its line of sight (LOS) and field of view (FOV) are defined by the Cassegrain telescope feeding light into the spectrograph. The extensible XRS boom cannot be employed in the ground-based full system optical test of the observatory due to the finite source focal length correction requirement. This, combined with errors introduced into the system at the boom mounting interfaces, and the boom’s deployment dispersion means that the XRS LOS is not predictable at the arcsec level based on information available before on-orbit deployment. Once the instrument is fully deployed on-orbit, its LOS can be compared to that of UVS. The XRS Alignment and Focus Mechanism (XAFM) will be used to adjust for any XRS pointing errors associated with channel alignment, boom mounting and deployment, XRS to Science Instrument Package Adapter Deck (SIPAD, i.e., the instrument mounting plate on the spacecraft) mounting issues, or distortion in the SIPAD itself.

In the next sections, we outline the nature of the Arcus Probe instrument internal alignment and the details of the error contributions to the XRS to UVS coalignment.

\subsection{Definition of FOV and LOS for XRS and UVS}

\subsubsection{Defining UVS LOS and FOV}
UVS is fed by a Cassegrain telescope. Its FOV is defined by the projection of the spectrograph slit at the UVS focal plane.  This angular projection is determined by the physical width of the slit and the focal length of the instrument. UVS's line of sight is a vector toward the source that forms an image at the center of the FOV. The UVS FOV has an angular projection on the sky of 10'' x 360'' with a 30'' x 30'' square at the center used for point source observations and target acquisition. UVS has a movable slit plate. Therefore, the line of sight and the FOV can be moved either by sliding the slit plate, or by tilting the entire instrument.

\subsubsection{Defining XRS LOS and FOV}

Before defining the LOS and FOV in detail, we define the following terms:

\begin{description}
    \item[Silicon Pore Optics Mirror Modules (SPO MM)] The X-ray collectors (essentially telescopes) that capture and focus X-ray light.
    \item[Grating facets] The individual mounted ~30 mm by ~30 mm critical-angle transmission (CAT) gratings that diffract the light that the SPO MMs capture and focus.
    \item[SPO Petal] The assembly that includes the structure and 40 silicon pore optics mirror modules (SPO MM) mounted on it and coaligned to produce a single image.
    \item[Grating Petal Assembly] The assembly that includes the structure and 216 grating facets mounted on it and coaligned to produce a single spectrum from the focused light transmitted by its associated SPO Petal.
    \item[Channel Optics Package (COP)] The combination of a single SPO petal and a single grating petal.
    \item[Channel] The combination of a single COP and the detectors.
    \item[X-ray Alignment and Focus Mechanism (XAFM)] a mechanism that can orient the Optics Platform (OP) in six degrees of freedom.
    \item[Optics Platform (OP)] the assembly mounted through the XAFM to the Forward Assembly (FA) that includes all four COPs and the structure that combines them.
    \item[Detectors] There are two detector assemblies, each consisting of an 8 by 1 detector array. Each channel uses some of both arrays. For a given channel, its zero-order image and low diffraction orders fall on one detector array, higher diffraction orders on the other. For the discussion of the co-alignment this operational detail is ignored.
\end{description}

XRS is a grazing incidence slitless soft-X-ray spectrometer that consists of four semi-independent optical channels feeding two cameras. The LOS of the XRS is a best fit of the lines of sight of the individual channels, which in turn are each the result of the relative position of the COPs and their zero-order images on the detector. Unlike the UVS, the FOV and LOS are independent for two reasons. First, the long thin pores in the SPO MMs vignette off-axis photons which reduces the effective area of the XRS. We therefore define the FOV based on minimizing this vignetting. Second, the COPs behave like a thin lens, so changing the orientation of the COPs does not affect where a source appears in the focal plane to first order. Thus changes in the orientation of the COPs do not change the instrument's LOS appreciably. As a consequence of these attributes the axis of minimum vignetting does not necessarily align with LOS of the XRS. In practice, the XAFM is used to ensure that the line of the XRS is pointed where it needs to be and the axis of minimum vignetting, and therefore the FOV, is aligned with the LOS.

We will define XRS pointing at the single channel level, then discuss the effects of generalizing that definition to the combined LOS and FOV.

\subsubsection{LOS of an individual X-ray channel}
A channel's line of sight is defined by a vector starting at the zero-order image, extending through the common center of curvature SPO MMs in the COP all the way to the celestial source (Figure 2).

\begin{figure}
    \begin{center}
      \includegraphics[width=0.48\textwidth]{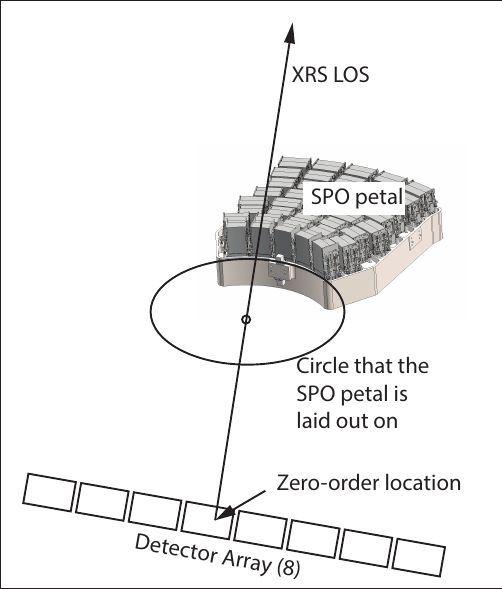}
    \end{center}
    \caption{The XRS LOS is given by a vector running from the zero-order image through the center of the SPO alignment.\label{fig:2}}
\end{figure}

\subsubsection{The XRS field of view}
\begin{figure}
    \begin{center}
      \includegraphics[width=0.48\textwidth]{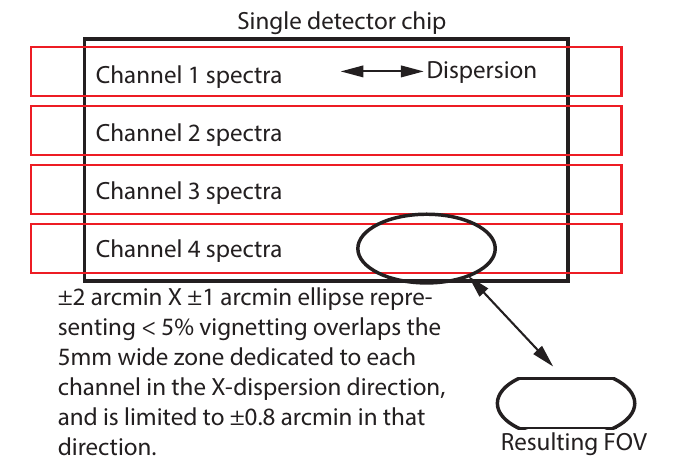}
    \end{center}
    \caption{The FOV of an XRS channel defined\label{fig:3}}
\end{figure}
Two factors combine to define XRS's FOV: the detector size (or the portion of the detector allocated to the channel), and vignetting caused in the long narrow pores in the SPO MMs. The detectors extend so far in the dispersion direction that they effectively have no role in defining the instrument FOV in that direction. Instead SPO induced vignetting in the dispersion direction, required to be limited to $< 5$\%, limits the FOV to $\pm2$~arcmin in that direction (See Figure~\ref{fig:3}). In the cross-dispersion direction, which aligns with the short axis of the detector, the 5\% vignetting requirement and the portion of a detector dedicated to a single channel play a nearly equal role in limiting the FOV, ($\pm0.98$~arcmin for vignetting versus $\pm0.8$~arcmin for the channel width). Since the channel width is more restrictive, the FOV in the cross-dispersion direction is limited to $\pm0.8$~arcmin, and the ellipse shape that results from a FOV defined solely by vignetting is trimmed on both sides.

\subsubsection{XRS combined LOS and FOV}
Each of the four channels that make up the XRS has its own LOS and FOV. During the process of mounting and aligning COPs into the optics platform (OP), the COPs will be oriented such that their vignetting axes are aligned with each other to within allowable errors. In addition, the COPs will be linearly positioned such that the resulting individual channel LOS are nearly co-linear (again, to within the allocated error budget). With the channels' LOS and FOVs aligned, the OP can be moved linearly in the plane of its structure to re-point the LOS of all 4 channels at once, and the OP can be rotated about vectors in the plane of its structure to orient all of the FOVs at once.

\subsubsection{Determining the pointing and alignment of the XRT}
If the XRS instrument is rotated as a whole, then the orientation of the LOS, and the orientation of the FOV will change, though their relationship to each other will remain invariant. However, when moving the OP by itself with the XAFM, adjustments to the LOS and FOV defined by vignetting can be made independently i.e., we can change one without affecting the other. The XRS line of sight  is changed by translating the OP along vectors in the OP plane, while the center of the FOV can be moved by rotating the OP about vectors in the OP plane. When the vignetting-defined FOV is centered on the XRS LOS, the OP is internally aligned and pointed unambiguously along the LOS.

\begin{figure}
    \begin{center}
      \includegraphics[width=\textwidth]{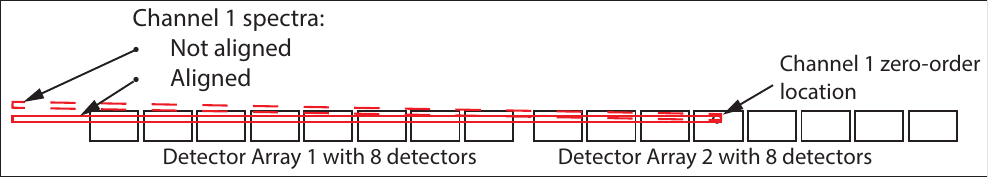}
    \end{center}
    \caption{Alignment of the XRS spectra and the detector arrays. Note that the detector spacing is not shown to scale - the distance between the two detector arrays is reduced for display purposes.\label{fig:4}}
\end{figure}

There are two other adjustments required to bring the XRS channels into full alignment. First, the line on which the diffracted spectral image forms needs to align with the long direction of the detector arrays, otherwise parts of it will fall off the detectors (See Figure~\ref{fig:4}). Adjusting this requires that the OP be rolled about the LOS axis. Second, the instrument needs to be focused. This is done by translating the OP along the LOS axis.

\subsection{Aligning XRS and UVS}
\subsubsection{What does alignment look like}

\begin{figure}
    \begin{center}
      \includegraphics[width=0.4\textwidth]{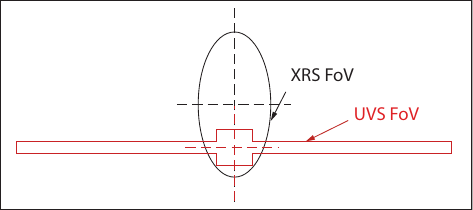}
    \end{center}
    \caption{XRS and UVS with overlapping fields of view\label{fig:5}}
\end{figure}

Since Arcus' science requirements are defined by point source observations, achieving alignment of both instruments requires that at least 5 arcsec by 5 arcsec of the sky is covered by both instruments simultaneously. However, Arcus Probe is considered to be in alignment, and will be maintained as such throughout the mission, when the XRS and UVS FOVs fully overlap (see Figure~\ref{fig:5}).

\subsubsection{Initial Alignment During Commissioning}
Once the Arcus Probe Observatory reaches orbit and both XRS and UVS are fully deployed, the spacecraft will be pointed toward a bright celestial source to gauge the co-alignment between the instruments and between UVS and the spacecraft.

As the UVS drives the pointing requirements of the Observatory, we select it as the pointed portion of the payload. This choice drives an aspect of the observatory design: the Observatory pointing references (three star trackers and an Inertial Measurement Unit (IMU)) are placed on the UVS structure to optimize UVS pointing. While the alignment between the Observatory Z-axis and the UVS line of sight is set on the ground, we allow for shifts during launch and/or imperfect ground calibration. If that happens, the resulting misalignment between UVS and the spacecraft will be adjusted by applying numerical offsets to the spacecraft pointing direction in the pointing control algorithm.  The misalignment between XRS and UVS cannot be corrected by an observatory pointing adjustment. Instead,  re-alignment of the two instruments must be accomplished by physically moving the FOV of one of the instruments through an internal optical adjustment.

\begin{figure}{r}
    \begin{center}
      \includegraphics[width=0.5\textwidth]{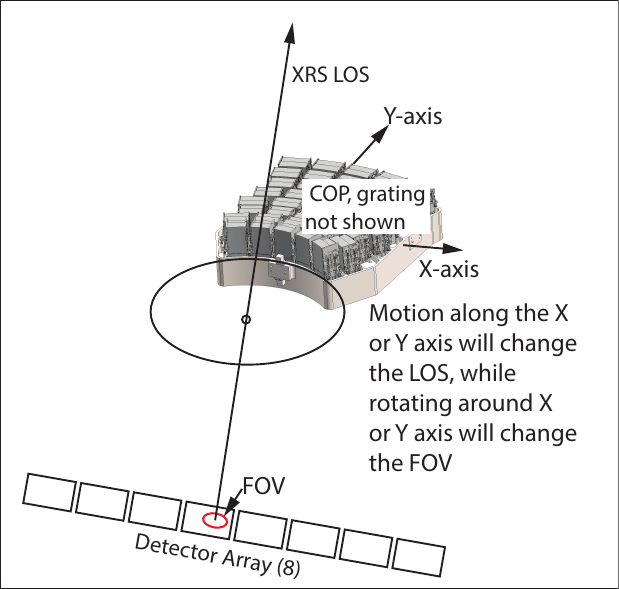}
    \end{center}
    \caption{COP (and OP) motions that affect XRS LOS and FOV alignment\label{fig:6}}
\end{figure}

If there is an offset between the XRS and UVS, we will use the XAFM to adjust for it. The procedure for doing so is as follows: (1)~Arcus Probe is pointed toward a bright celestial source. The source is centered in the UVS slit, and thus the UVS LOS. Any variations between the spacecraft Z-axis and the UVS LOS are corrected in the spacecraft pointing offsets. (2)~The location of the image in the XRS focal plane is compared to the desired location. The XAFM will be used to translate the OP until the XRS LOS and the UVS LOS  are aligned (Figure~\ref{fig:6}). (3)~The OP will be tilted about the X and Y axes to minimize vignetting at the new LOS. As the XRS is a thin lens, these tilts do not affect the LOS but only the FOV. The best tilt angles will be determined post-facto by comparing flux levels at each position. Finally, the OP will be rolled to ensure spectral alignment with the long axis of the detector arrays, and translated in the Z-axis to optimize focus.

For the remainder of the mission, the relative alignment of the two instruments will be tracked, correlating changes in their relative alignment with the Observatory's orientation to the Sun and Earth to ensure co-alignment between the two instruments throughout the science mission. Given each instrument's large FOV, the benign thermal environment, active thermal control, and selected materials, we do not anticipate needing to adjust the XRS to UVS relative alignment again after commissioning is complete, though nothing about the alignment process proposed here would prohibit us from doing so.

\subsubsection{Types of errors}
There are four types of XRS internal errors:
\begin{itemize}
    \item Errors that change pointing: linear motions in X and Y of either the COPs, or the OP;
    \item Errors that change the angle of minimum vignetting: rotation motions around X or Y of either the COPs or the OP;
    \item Errors that misalign the XRS spectra and the detectors: rotations of the OP around the Z axis;
    \item Errors in focus: linear motion of the OP along  Z axis.
\end{itemize}
The first two errors affect co-alignment with the UVS, while the second two are the result of buildup and initial deployment. All four sources of internal error are controlled with the XAFM (performance shown in Table~\ref{tab:ranges}).

Once the XRS has been deployed and aligned with the UVS, the main error sources in co-alignment are deflections in the SIPAD that both the UVS and the XRS are mounted to, and errors in the boom induced by asymmetric temperatures in the longerons.

\subsubsection{Error budgets}
\paragraph{Errors during the buildup of the OP}
Current best estimates (CBE) for the single COP errors are listed in Table~\ref{tab:singleCOP}.

\begin{table}
    \caption{Current best estimate (CBE) performance and alignment effects of the placement of single COPs. Cell colors indicate misalingments that are connected, e.g., 200~micron displacement of the OP (or a COP) yields a 3 arcsec LOS error. Bold fond numbers are derived from RSS combining individual contributions.\label{tab:singleCOP}}
    \vspace{2mm}
    \centering
\begin{tabular}{|l|rrr|rrr|}
    \hline
     & $\Delta x$ & $\Delta y$ & $\Delta z$ & rot $\Delta x$ & rot $\Delta y$ & rot $\Delta z$\\
     & [$\mu$m] & [$\mu$m] & [$\mu$m] & [arcsec] & [arcsec] & [arcsec] \\
\hline\hline
\textbf{Single COPs} & & & & & & \\
~Knowledge of the COP center & 10 & 10 & 10 & 1 & 1 & 1 \\
~COP cube as a reference     & 10 & 10 & 10 & 2 & 2 & 2 \\
~Placement of each COP       & 200 & 200 & 400 & 10 & 10 & 10 \\
\hline
\textbf{Errors in COP placement} & \cellcolor{lime} \textbf{200} & \cellcolor{lightgray}\textbf{200} & \textbf{400} & \textbf{10} & \textbf{10} & \textbf{10} \\
\textbf{Zone of FOV error between COPs} & & & & \cellcolor{lightgray} \textbf{3} & \cellcolor{lime} \textbf{3} & \\
\hline
\end{tabular}
\end{table}

There are three main error sources at the COP level:
\begin{itemize}
    \item \textbf{Knowledge of the COP configuration:}
This is an understanding of where the COP is pointed and where the axis of minimum vignetting is. These are easy to determine since they are set in the construction of the SPO petal itself and later confirmed at PANTER (COP 1) and the XRCF (OP).
\item \textbf{Reference cube placement:}
Once we know the COP orientation, we place a cube on it to witness that orientation. The placement has an inherent set of tolerances that are understood and well-determined from years of similar alignment and placement (e.g., SDO/AIA, IRIS, multiple campaigns of the HiC rocket mission, etc.).
\item \textbf{Physically placing the COPs in the OP structure:}
Actually, placing the hardware has the largest associated uncertainty. We allocate 200 microns for the X and Y axes, and 400 microns for the Z axis (focus). The linear placement error of the COPs can be reduced, if needed, but has little effect on the overall system alignment.
\end{itemize}

The final term in Table \ref{tab:singleCOP}, Zone of FOV error between COPs, indicates the resulting misalignment of channel line of sight to the individual COPs. The practical effect of this is to reduce the effective size of the OP FOV (see overlapping ellipses on the right side of Figure 7).

\paragraph{Internal Instrument alignment errors}
CBEs for the internal instrument alignment errors are listed in table~\ref{tab:2}.

\begin{table}
    \caption{CBE for internal XRS alignment errors that affect alignment with the UVS\label{tab:2}}
    \centering
\begin{tabular}{|l|rrr|rrr|}
    \hline
     & $\Delta x$ & $\Delta y$ & $\Delta z$ & rot $\Delta x$ & rot $\Delta y$ & rot $\Delta z$\\
     & [$\mu$m] & [$\mu$m] & [$\mu$m] & [arcsec] & [arcsec] & [arcsec] \\
\hline\hline
\textbf{The OP (overall alignment)} & & & & & & \\
~RA Cube location & 10 & 10 & 10 & 5 & 5 & 5 \\
~RA mounting (RA linear errors do not affect alignment) & N/A &  N/A &  N/A & 5 & 5 & 5 \\
~Boom to RA mounting & 10 & 10 & 10 & 5 & 5 & 5 \\
~FA Cube Location & 10 & 10 & 10 & 5 & 5 & 5 \\
~Boom to FA mounting & 10 & 10 & 10 & 5 & 5 & 5 \\
~Boom opening dispersion & 50 & 7 & 3 & 7 & 7 & 60 \\

\hline
\textbf{Error from XRS integration of OP} &  \textbf{327} &  \textbf{297} & \textbf{39} & \textbf{13} & \textbf{13} & \textbf{61} \\
\hline
\textbf{Combined COP and OP error} &  \textbf{384} &  \textbf{446} & \textbf{402} & \textbf{17} & \textbf{17} & \textbf{62} \\
\hline
\end{tabular}
\end{table}

Internal instrument alignment errors consist of a set of errors associated with putting the overall XRS instrument together. They include:
\begin{itemize}
    \item \textbf{RA Cube location.} This is an error that results from the difference between the alignment cube's true orientation and our measurements of its orientation. It also embodies the errors in our understanding of where the detectors are in the RA and errors in conveying that understanding to their cube mounting location and orientation.
    \item \textbf{RA mounting.} This error results from mounting the RA to the instrument support plate. There is no linear component to this error since translating the RA has no effect on the co-alignment of the XRS and UVS. This error results in a pointing error between the XRS and UVS.
    \item \textbf{Boom to RA mounting.} Angular errors in this term have a significant impact, as they have the effect of the moving the OP linearly over a \~11 m lever arm.
    \item \textbf{FA cube location.} This error is similar to the RA cube location error.
    \item \textbf{Boom to FA mounting.} The effects of the angular errors have a similar effect as the those in the RA mounting row, though the mount arm is shorter (\~1.5m).
    \item \textbf{Boom opening dispersion.} There is a range of positions for the position of the front of the boom assuming a fixed position of the rear of the boom. We will select the mean of each set of ground based measurements associated with a given degree of freedom (X, Y, etc.) as the boom position to use when calculating the attachment orientation between each end of the boom and its attachment point. The boom will open somewhere is the measured range, but most likely away from the mean. This is treated as an alignment error. These errors have a similar effect to the FA mounting errors.
\end{itemize}

\paragraph{Combination of lower-level errors and external sources of misalignment}
The final set of errors show the combined effect of those at lower levels, external sources of error, and the effect these have on internal alignment and co-alignment between XRS and UVS. The CBE for those errors are listed in table~\ref{tab:err_combined}.
\begin{table}
    \caption{Final alignment errors as they affect XRS to UVS Co-alignment at commissioning\label{tab:err_combined}}
    \centering
\begin{tabular}{|l|rrr|rrr|}
    \hline
     & $\Delta x$ & $\Delta y$ & $\Delta z$ & rot $\Delta x$ & rot $\Delta y$ & rot $\Delta z$\\
     & [$\mu$m] & [$\mu$m] & [$\mu$m] & [arcsec] & [arcsec] & [arcsec] \\
\hline\hline
\textbf{Mounting plate etc.\ alignment errors} &  0 &  0 &  0 & 5 & 22 & 0 \\
\textbf{Total error in spectral image position} & \cellcolor{pink} \textbf{384} & \cellcolor{gray} \textbf{446} & \textbf{402} & & & \\
\textbf{Overall OP initial internal pointing error}  &  &  &   & \cellcolor{gray} \textbf{10} & \cellcolor{pink} \textbf{10} &  \\
\textbf{Pointing error of the XRS wrt UVS}      & & & &  \textbf{12} &  \textbf{24} & \\
\textbf{COP FOV offset, and spectrum rotation} &  & & &  &   17 &   62 \\
\hline
\end{tabular}
\end{table}

\begin{itemize}
    \item \textbf{Mounting plate etc. alignment errors} These are errors resulting from outside sources. For example, this term includes flexure in the SIPAD that both XRS and UVS are mounted to. The SIPAD has a predicted performance of 18.1 arcsec 3$\sigma$ around the Y axis (aligned with the UVS slit and the cross-axis direction of XRS). There is an additional error allowance of 12 arcsec 3$\sigma$ for other sources of rotY error that are added via a quadrature sum to yield the full allocation of 22 arcsec. We anticipate that there will be very little need for later alignment adjustments.
    \item \textbf{Total error in spectral image position} This is a combination of all error sources internal to the XRS that will result in the translation of the spectral image (X and Y) and therefore can be treated as a change in the line of sight. Error in the Z direction is a focus error.
    \item \textbf{Overall OP initial internal pointing error} This is the effect that OP linear position errors have on the line of sight of the XRS. They are calculated by scaling $\Delta x$ and $\Delta y$ by the focal length (12m) and converting to arcsec.
    \item \textbf{Pointing error of the XRS WRT UVS} This is the combination of the pointing error induced by the mounting plate combined with the internally induced pointing error. The plate plus other external sources dominate.
    \item \textbf{COP FOV offset, and spectrum rotation} RotX and rotY enumerate the offset between individual COPs and represents the reduction in FOV that results (Figure~\ref{fig:7}, righthand side). RotZ enumerates the effect shown in Figure~\ref{fig:4}.
\end{itemize}

\subsection{Co-aligning the XRS to the UVS at Commissioning}
The co-alignment of the Arcus Probe instruments involves addressing several factors:
\begin{itemize}
    \item Errors in the alignment of the XRS at commissioning;
    \item Errors in the alignment of the UVS at commissioning;
    \item Errors in the SIPAD and instrument mounting at commissioning;
    \item Variations in all of the above over time after commissioning;
\end{itemize}

\begin{figure}
    \begin{center}
      \includegraphics[width=\textwidth]{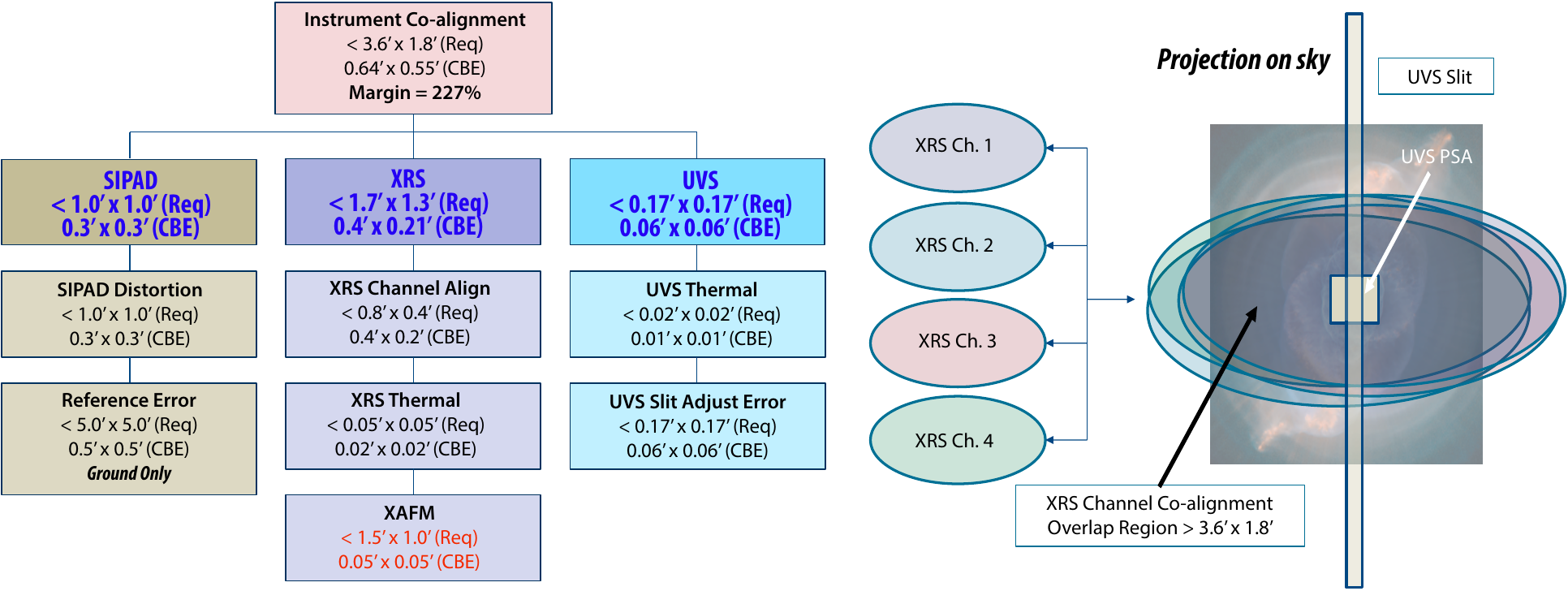}
    \end{center}
    \caption{Overall Arcus Probe Instrument Alignment (COP FOV simplified to ellipses)\label{fig:7}}
\end{figure}

These issues are dealt with in the same way: The XAFM brings the XRS and the UVS into alignment. Above, we assumed that the UVS alignment errors dropped out to ease the discussion, but to determine the required XAFM range we need to track UVS errors as well. Figure~\ref{fig:7} ties in the expected pointing stability of the UVS, XRS, and SIPAD, rolling up the numbers that we developed above with the allocated SIPAD performance and the predicted UVS performance. Table~\ref{tab:ranges} shows the design adjustment range of the XAFM, the required range, and the margin.

\begin{table}
    \caption{Range, resolution, and margin of the XAFM\label{tab:ranges}}
    \centering
\begin{tabular}{|l|rr|rr|}
    \hline
 & Range & Margin on Range & Resolution & Margin on resolution \\
    \hline\hline
X & 15 mm & 29 X & 1 $\mu$m & 399 X \\
Y & 15 mm & 29 X & 1 $\mu$m & 399 X \\
Z & 15 mm & 14 X & 1 $\mu$m & 199 X \\
rotX & 229.2 arcmin & 56.3 X & 0.0034 arcmin & 47.5 x\\
rotY & 229.2 arcmin & 56.3 X & 0.0034 arcmin & 47.5 x\\
rotZ & 114.6 arcmin & 27.6 X & 0.0069 arcmin & 144.4 x\\
\hline
\end{tabular}
\end{table}

\section{RAY-TRACING}
\label{sect:raytracing}
\subsection{Optical Design}
\label{sect:design}

\begin{figure} [ht]
    \begin{center}
    \begin{tabular}{c} 
    \includegraphics[width=\textwidth]{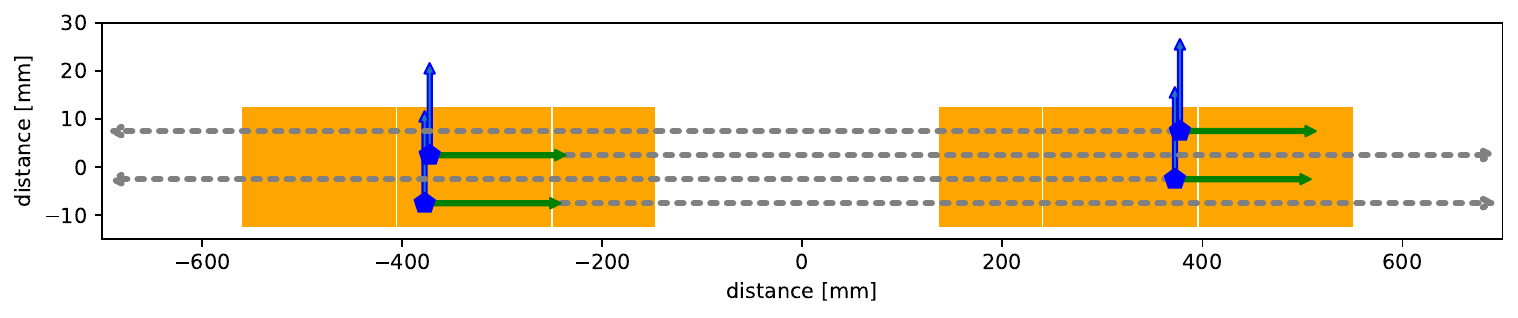}
    \end{tabular}
    \end{center}
    \caption {\label{fig:geometry_camera}
    Layout of the CCDs (orange rectangles) in the focal plane. Given the resolution of the figure, most chip gaps are not visible. Also note, that the x and the y axes are scaled differently (Figure~\ref{fig:4} shows correct aspect of the CCDS, but not the correct gap between the two detector arrays). Two cameras with eight CCDs each are located in the focal plane. In this plot, 0 is the geometric center of all the Rowland tori involved. Note that the cameras are intentionally not symmetric to 0.
    The position of the four optical axes for the four channels is marked with a blue pentagon.  The dispersion ($x$ - green horizontal arrows) and the cross-dispersion ($y$ - blue vertical arrows) of the different spectra are parallel to each other and point in the same direction. Two channels disperse left-to-right, and two right-to-left.
    }
\end{figure}

Each of the four channels of the XRS has a separate zero-order position and disperses the light at a different location on the detector. This has two main advantages: there is no need to align all COPs to the same zero-order position within the width of the point-spread function (PSF, less than 1.5 arcsec in dispersion direction), and any wavelength of light that falls in a chip gap for one channel will be observable in the others; see Figure~\ref{fig:geometry_camera}. At the same time, this layout allows us to use the same detectors for all four channels, which reduces cost and complexity. More details of the optical layout are given in our previous ray-tracing papers\cite{10.1117/12.2273011,10.1117/12.2312678}.

In each Grating Petal Assembly, CAT gratings are mounted in a Rowland-torus
design\cite{Beuermann:78}, where the centers of the gratings are arranged on the surface of a torus. The gratings are blazed, and thus the light is diffracted mostly to one side; most of the dispersed photons are found around twice the blaze angle in an area called the ``blaze peak.'' The two cameras are positioned to capture the zeroth order on one side and the blaze peak on the other with a gap in between where so little signal is found that it is not useful to capture it with CCDs.

The layout of four channels with different zeroth orders and dispersed spectra utilizing the same detectors is possible due to a ``tilted rowland torus'' geometry, where the optical axis of a COP does not intersect with the center of the Rowland torus, but is offset\cite{10.1117/12.856482,10.1117/12.2273011}. In this geometry, the zeroth order and the center of the blaze peak are both in the focal plane. The four channels are arranged in pairs. Two of them have the zeroth order on the left in Figure~\ref{fig:geometry_camera} and disperse to the right, two of them have the zeroth order on the right and disperse towards the left. The pair of channels dispersing in the same direction set the CAT gratings on very similar Rowland tori, which are offset from each other in cross-dispersion direction by 10~mm so that the signal is clearly separated on the CCDs. They are also offset from each other in dispersion direction by 5~mm, to ensure that the same wavelength seen in different channels falls onto different x coordinates on the CCDs and thus a wavelength that happens to be lost in a chip gap between two CCDs in one channel can be observed in the other channel.

\subsection{Ray-trace inputs}
We use the ray-tracing code MARXS\cite{marxs1.1,2017AJ....154..243G}, a Python code developed by us under an open source license. MARXS is available at \url{https://github.com/chandra-marx/marxs}; simulations in this article are done with a development version with git hash f4b68dce0957 to include Arcus specific code. MARXS has been used for several mission proposals and is backed by hundreds of units tests and verified against laboratory works and on-orbit Chandra observations where feasible. MARXS does not calculate material properties such as reflectivities or grating efficiencies; instead they are read in from tables that can be based on laboratory data or simulations with other programs.

\subsubsection{SPOs}
\begin{figure}
    \centering
    \includegraphics[height=8cm]{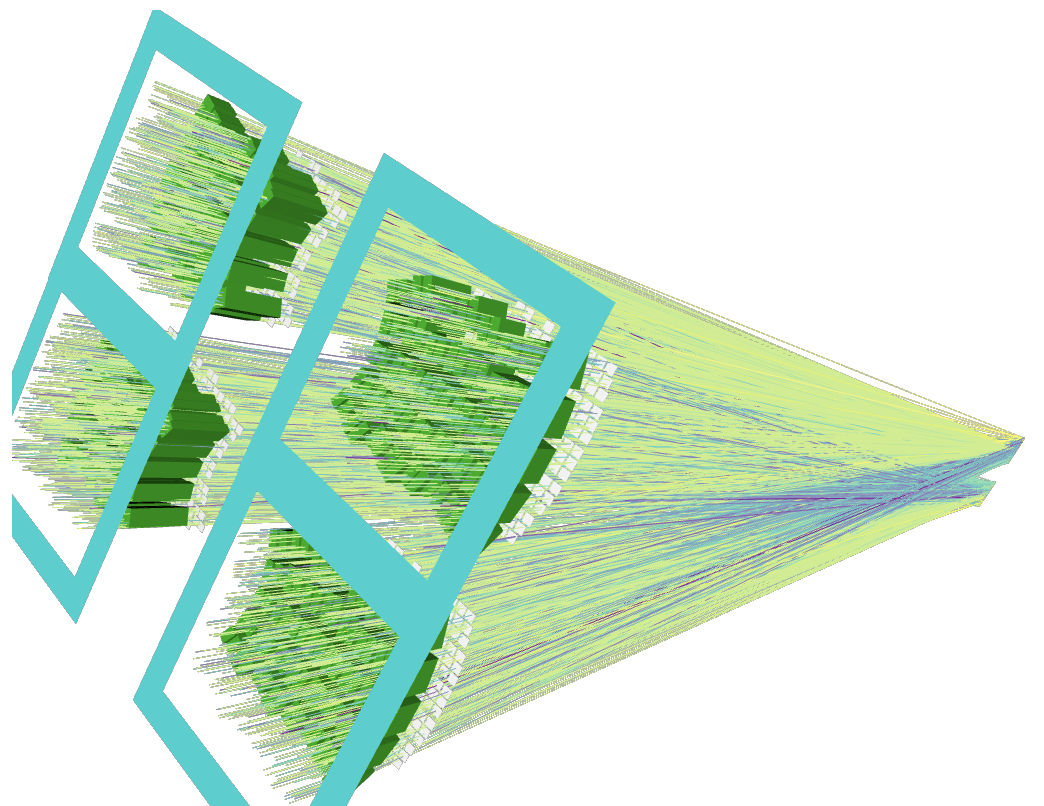}
    \caption {\label{fig:arcus}
    Arcus to scale. On the left are the four channels with SPOs (green) and CAT gratings (white). The CCDs are in the background, where the lines converge. Lines show ray-traces. Different colors indicate different diffraction orders. Only rays that reach the detector are shown.
}
\end{figure}

The SPOs are a technology developed and matured for Athena. They provide a large effective area at a relatively low weight and cost\cite{10.1117/12.2188988,10.1117/12.2599339} - see also Girou et al.\ in this volume for the most recent update. In Arcus, the SPO petals use ``sub-aperturing,'' which means that the SPOs do not cover a full circle, but only a narrow wedge. That provides Arcus with an asymmetric PSF that is only about 1.5\ arcsec wide in the dispersion direction. It also means that the individual petals where the SPOs are mounted are not circular, but roughly rectangular, and four separate panels (one per optical channel) can fit into the FA; see Figure~\ref{fig:arcus} for a rendering. Our SPO model is simplified and does not take into account vignetting, which has been discussed in great detail in the previous section; but it does consider the geometric opening area (incl.\ ribs) and angle-dependent reflectivity to predict the effective area.

\subsubsection{CAT gratings}
CAT gratings are manufactured at the Space Nanotechnology Laboratory. Progress on the development and manufacturing of the CAT gratings is given by Heilmann et al.\ in this volume and a series of papers stretching back many years\cite{10.1117/12.926827,2022ApJ...934..171H}. For our ray-trace simulations, we use CAT grating efficiences that are predicted from numerical simulations and that match the efficiences measured for CAT gratings in the laboratory. The numbers include the effect of cross-dispersion from L1 support structures, dispersion and blockage by the hexagonal L2 supports and any coverage by grating frames.

\subsection{Effective area and resolving power}
The main performance characteristics of Arcus are the effective area and the resolving power of the extracted spectra. We run simulations on a wavelength grid from 1.5 to 60 \AA{} in steps of 0.15~\AA{}. Each simulation is run with 100,000 rays. MARXS tracks the probability of survival for each ray; for example when a ray passes a filter that transmits 15\% of the photons at that energy, the probability of a ray passing that filter is multiplied by $0.15$. In the end, the sum of the probabilities of all photons hitting the detector divided by the total number of rays simulated for a given entrance aperture gives the effective area with much lower statistical noise than in codes where photons either pass or do not pass individual elements. The uncertainty on the Arcus performance is dominated by the uncertainty in the material properties such as SPO reflectivity or CAT grating efficiency. These systematic effects are difficult to quantify, but are probably of order 10-20\% or so in sum, while the statistical uncertainty for each run is below a few per cent.

\begin{figure}
    \centering
    \includegraphics[width=0.8\textwidth]{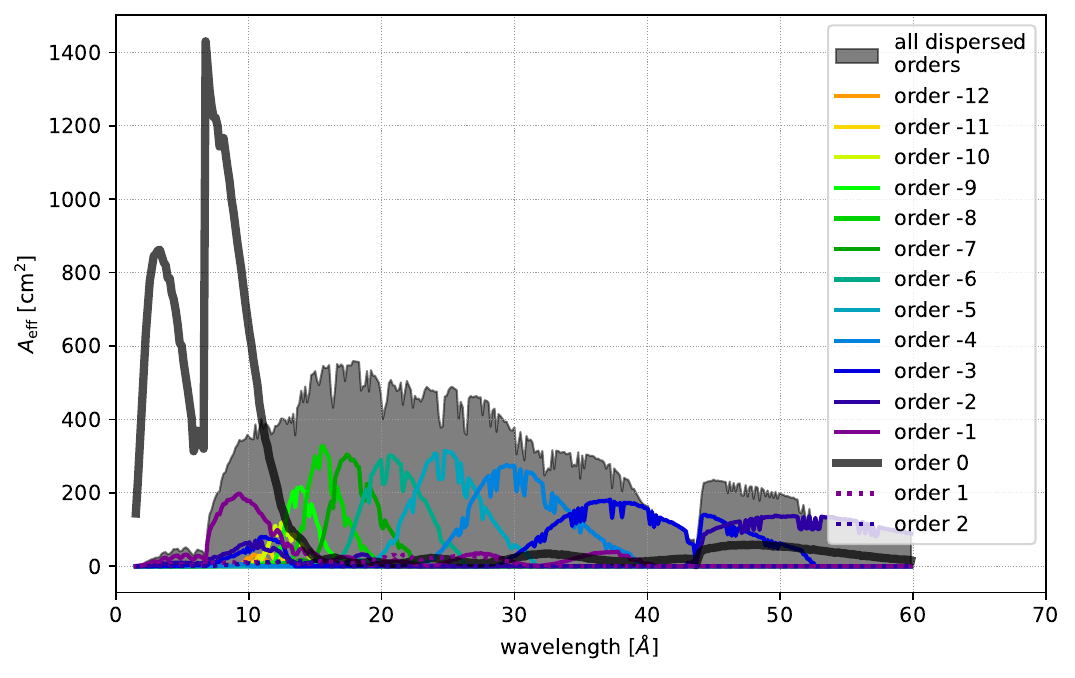}
    \caption {\label{fig:Aeff}
        Effective area for Arcus based on ray-trace simulations. The effective area of individual dispersion orders is shown. For most wavelength ranges, several dispersed orders contribute to the total effective area.
    }
\end{figure}
\begin{figure}
    \centering
    \includegraphics[width=0.8\textwidth]{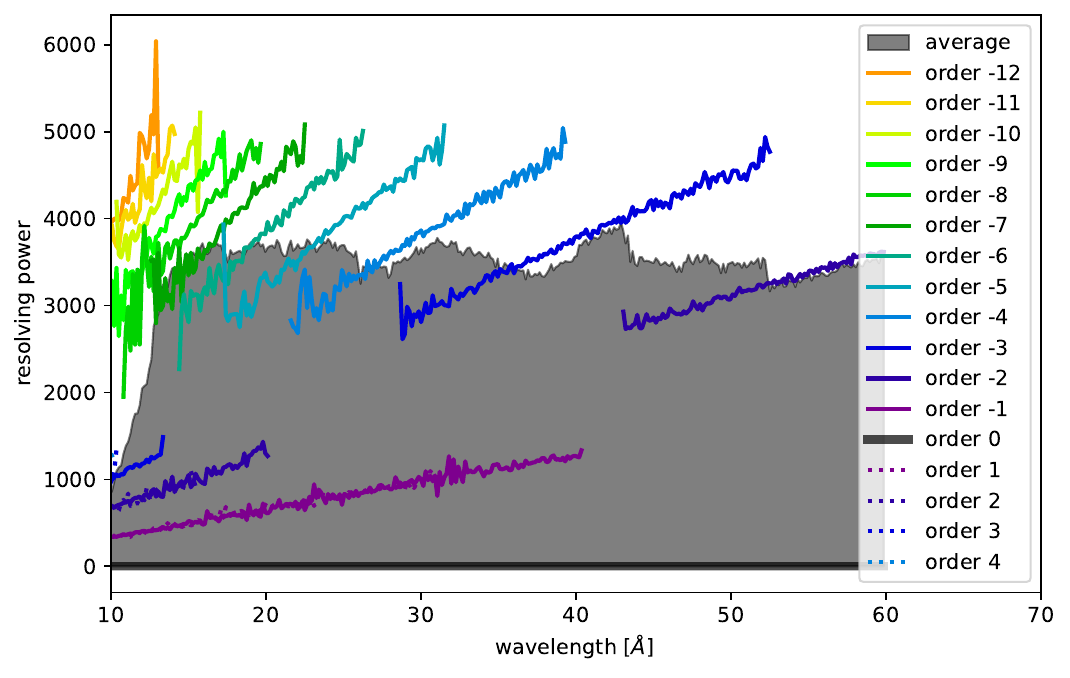}

    \caption {\label{fig:R}
    Resolving power $R$ for Arcus. Colored lines show $R$ per order, the gray area shows $R$ averaged over all diffracted orders contributing to that wavelength, where the average is calculated weighting each order by the effective area. For low orders, there are gaps in the colored lines. At low wavelength, those orders fall on the same camera as the zeroth order, e.g.\ for order -2 up to 20~\AA{}. In order -2 photons longer than 43~\AA{} are dispersed into the blaze peak and detected on the other camera.
    }
\end{figure}

The effective area for Arcus is shown in Figure~\ref{fig:Aeff} and the spectral resolving power $R$ per order and averaged over all orders that contribute at a particular wavelength (weighted by the effective area of each order at that wavelength) in Figure~\ref{fig:R}. The figure distinguishes between the photons seen in dispersion and in order zero, i.e.\ in direct light, where the only energy resolution is provided by the CCD. The effective area in direct light is high for high energies, in particular including the 6.7~keV (1.85~\AA{}) iron line, which is a crucial diagnostic in many astrophysical sources. Towards longer wavelengths, around 10~\AA{}, light in order 1 and -1, makes up the bulk of the signal with $R$ of a few hundred. From about 12~\AA{} to 60~\AA{} most photons are dispersed into the blaze peak on the set of detectors opposite of the zeroth order and Arcus achieves an $R$ around 3500, almost independent of wavelength. The resolving power increases with increasing distance between the location of the zeroth other and the dispersed signal. For a given order, that means that $R$ increases with wavelength. However, for most wavelengths, multiple orders contribute to the signal. As the wavelength increases, the order on the far side of the blaze peak becomes weaker, and orders on the near side stronger, such that the average $R$ is approximately constant.

\subsection{Alignment tolerances}
We use ray-traces to assess the effect of misalignments on the performance of Arcus. Science requirements put limits on the maximal allowable degradation of spectral resolving power and effective area, and engineering constraints determine how well e.g.\ individual SPOs can be aligned into petals, how well the petals can be aligned to the forward assembly, and how well the forward assembly can be aligned to the detector housing. In general, tighter tolerances require more work, time, and money. We thus need to understand how important each possible degree of freedom is to the total performance of the system to identify those where significant design and work needs to go into the alignment.

The simulations start from a perfectly aligned version of Arcus. Even this does not provide infinite resolving power, because the model includes pointing jitter, a limited PSF, some astigmatism inherent in the design, and finite sizes of CAT gratings and CCD detectors, which means that they deviate from the ideal Rowland geometry. A ray-trace is run with this design and spectral resolving power ($R$) and effective area ($A_\mathrm{eff}$) are calculated.

After running the baseline version, one element of Arcus is shifted in one degree of freedom, e.g.\ all CCDs are shifted by 1~mm in the dispersion direction. The ray-trace is repeated, again $R$ and $A_\mathrm{eff}$ are calculated, then all CCDs are shifted by 2~mm and so on. After testing out the parameter space in dispersion direction, the CCDs will be shifted in cross-dispersion direction. In this way, each element (for example the CCD array), will be misaligned by various amounts in one of 6 degrees of freedom (shift along $x$, $y$, $z$, and rotation around $x$, $y$, $z$).
In the Arcus coordinate system, the $z$-axis is parallel to the optical axis, the $x$-axis is the grating dispersion direction and the $y$-axis is the cross-dispersion direction. In general, rotations are not done around the origin of the coordinate system, but around the center of an element (e.g.\ the center of a SPO petal).

In the first stage, only one degree of freedom is changed at a time, because it is not computationally feasible to explore the entire parameter space. From those results, we identify where the alignment is easily (e.g.\ just from simple machining tolerances) much better than the requirements. In a second step, we can then run ray-traces where all degrees of freedom are varied according to the misalignment budget and thus check if the assumptions going into combining the misalignments in different degrees of freedom hold or if non-linear interactions degrade $R$ and $A_\mathrm{eff}$ more than expected. In this case, we would have to revise the misalignment budget appropriately.

To keep the computational load reasonable, we simulate only one channel of Arcus. Since spectra from each channel will be extracted separately and there is symmetry between the channels, most results apply equally to all channels. We discuss in the text the few instances where the channel symmetry does not apply.
Each simulation consists of 200000 photons and the same source photons are used for each series of ray-traces, e.g.\ all global CCD misalignments use the same photons.

\begin{figure} [ht]
    \begin{center}
    \begin{tabular}{c} 
    \includegraphics[width=0.7\textwidth]{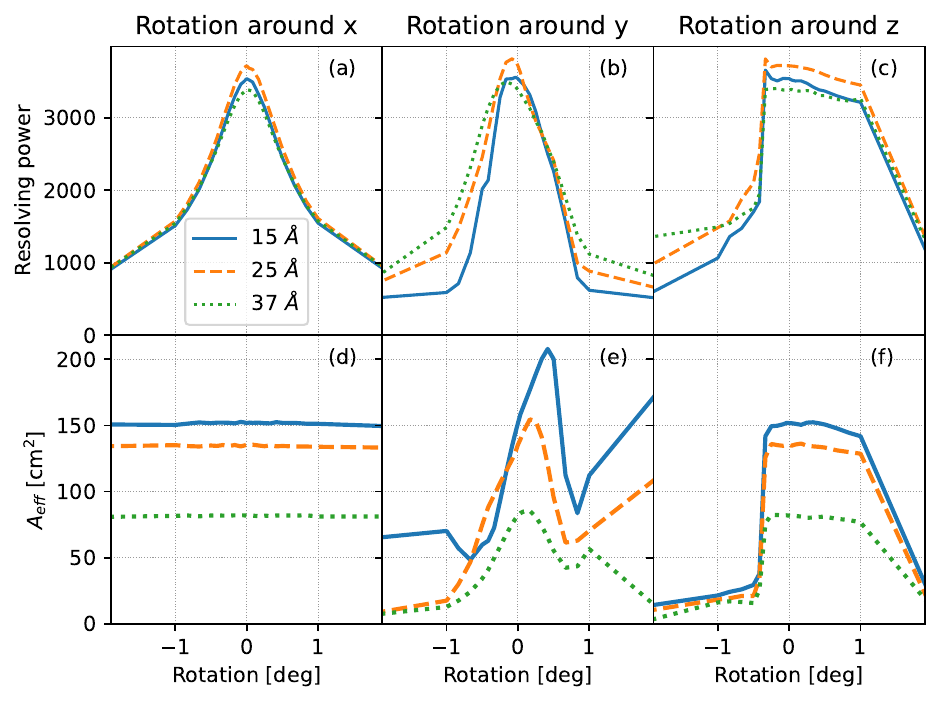}
    \end{tabular}
    \end{center}
    \caption {\label{fig:CAT_global}
    Effect of mislignment of the CAT grating petal for rotation around the center of the petal. (a-c): Resolving power for rotation around $x$, $y$, and $z$, respectively, (d-f): effective area for rotation around $x$, $y$, and $z$, respectively.
    }
\end{figure}
\begin{figure} [ht]
    \begin{center}
    \begin{tabular}{c} 
    \includegraphics[width=0.7\textwidth]{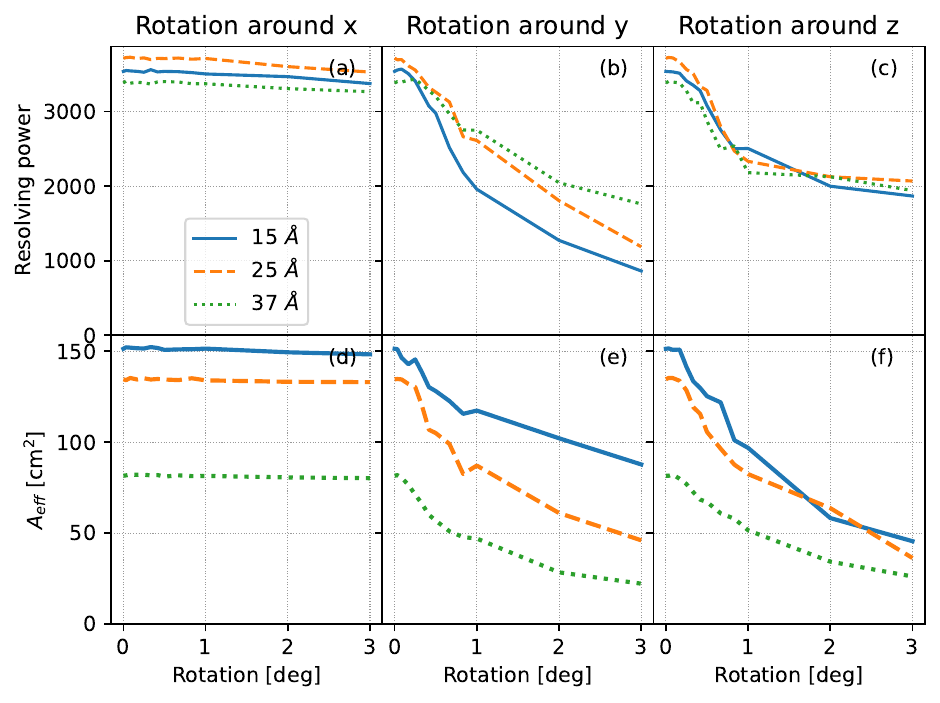}
    \end{tabular}
    \end{center}
    \caption {\label{fig:CAT_individual}
    Effect of mislignment of individual CAT gratings. The mislignment for each grating is drawn from a normal distribution, the $\sigma$ of that distribution varied from 0 to 3~degrees. (a-c): Resolving power for rotation around $x$, $y$, and $z$, respectively, (d-f): effective area for rotation around $x$, $y$, and $z$, respectively. Note that the $A_\mathrm{eff}$ reported is only for a single channel.
    }
\end{figure}
\begin{figure} [ht]
    \begin{center}
    \begin{tabular}{c} 
    \includegraphics[width=0.7\textwidth]{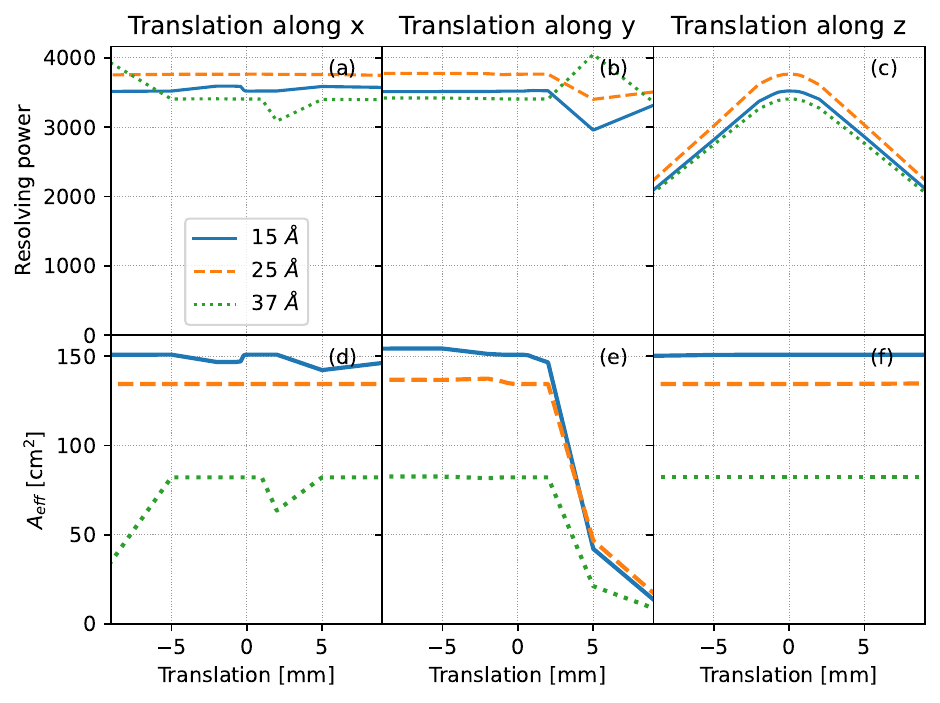}
    \end{tabular}
    \end{center}
    \caption {\label{fig:detector_global}
    Effect of mislignment the cameras with respect to the front assembly. (a-c): Resolving power for translation parallel to $x$, $y$, and $z$, respectively, (d-f): effective area for translation parallel to $x$, $y$, and $z$, respectively.
    }
\end{figure}

Figures~\ref{fig:CAT_global}, \ref{fig:CAT_individual}, and \ref{fig:detector_global} show some examples of the results. In the figures, $A_\mathrm{eff}$ is given summed over all dispersed orders that fall on a CCD (bottom row) and $R$ is the average resolving power, where the resolving power from individual orders is averaged weighted by the number of photons in that particular order (top row). Thus, it is possible, that $R$ in the plot increases with increasing misalignment if $A_\mathrm{eff}$ drops at the same time. This happens when an order with lower-than-average $R$ drops off the CCD (thus reducing the summed $A_\mathrm{eff}$ and increasing the average $R$). There is no scientific benefit from the apparently increased $R$ here - if one required a higher resolving power, the lower orders can be ignored in scientific analysis even if they fall on the CCD.

Figure~\ref{fig:CAT_global} shows the effect of a change in the CAT grating petal position, while the SPO petal and the cameras are fixed. Rotations around either $x$ or $y$ mean that the CAT gratings on one side move up, while the other side moves down, changing the path length of the diffracted photons. Those photons coming from the high CAT gratings travel further along the dispersion direction than those from the low gratings, thus causing the dispersed spot to smear out, which reduces $R$. These rotation need to be kept at the level of a few arcminutes. Rotation around $z$ changes the direction of the dispersed light and, for large angles, the dispersed orders miss the CCD.

Figure~\ref{fig:CAT_individual} examines the rotations of individual gratings. For each grating in the CAT petal, a different misalignment is drawn from a normal distribution. Rotations around $x$ (the dispersion direction, see panels a and d in the figure) have little effect on $R$. Since the facets have finite size, their edges differ from the Rowland torus by a little already, and adding a little more rotation does not change much. Rotations in $y$ quickly reduce $A_\mathrm{eff}$ though, because the CCDs are placed for a certain blaze peak. Rotating the CAT gratings shifts the blaze peak and photons miss the CCDs. Rotation around $z$ causes more and more signal to miss the detector. There is no sharp cut-off, because the CAT gratings in the simulation have a distribution of rotation angles, and the larger the Gaussian $\sigma$ is, the more CAT gratings will send their dispersed photons to positions where they cannot be detected. From this figure, we can determine that the alignment tolerance for rotation around $x$ can be large, while the other two directions are of order a few arcminutes to prevent significant degradation of the Arcus performance.

Figure~\ref{fig:detector_global} shows simulations for translating the detector with respect to the forward assembly (SPOs and CAT gratings). The most important degree of freedom is a change in focus (panel c). A shift along $y$ has no impact, as long as it is small enough to keep the dispersed spectrum on the CCDs. For the particular channel simulated here the spectrum drops off the CCDs for $y$ shifts for about -15~mm on one side and about +5~mm on the other side. A shift of 5~mm or more will drop at least one channel of the detector.

The curves for changing $R$ with shift in $x$ show some up and down when an order hits a chip gap (panel a and d). For example, two orders contribute to the signal in the curve for 37~\AA{} photons. At +2 mm one of them hits a chip gap, causing a drop in $A_\mathrm{eff}$ and also in $R$ ($R$ is averaged over all contributing orders, but only one order, which happens to have a lower $R$, is detected at this position). Note that chip gaps are inevitable. There will always be some wavelengths in a chip gap; however, the overall performance of the instrument, which is averaged over a range of wavelengths, is unaffected by this. While, in principle, a shift along $x$ does matter in principle since the focal plane is curved, we find that shifts up to a few mm have little impact, as if the CCDs move in the dispersion direction, the spectrum will be only slightly out of focus. As described above, the Arcus design takes advantage of this to mitigate the effect of chip gaps by offsetting the different channels by a few mm in dispersion direction so that no two spectra have chip gaps at the same wavelength.

We run simulations for rotations and translations for each possible mechanical misalignment in Arcus, as well as for a few other parameters such as the SPO mirror PSF, the repeatability of the CAT grating period, the surface flatness of the CAT gratings etc, but it is beyond the scope of this article to show each result in a figure. Inspecting all results, and considering how easy or hard it is to improve alignments in each degree of freedom, we set the alignment tolerances listed in Table~\ref{tab:tolerances}.

\begin{table}
    \caption{Arcus alignment tolerances ($1\sigma$)\label{tab:tolerances}}
    \centering
    \begin{tabular}{l|cccccc}
        \hline\hline
    alignment & trans x & trans y & trans z & rot x & rot y & rot z \\
     & $\mathrm{mm}$ & $\mathrm{mm}$ & $\mathrm{mm}$ & $\mathrm{{}^{\prime\prime}}$ & $\mathrm{{}^{\prime\prime}}$ & $\mathrm{{}^{\prime\prime}}$ \\
     \hline
    individual SPO in petal & 0.007 & 0.033 & 0.167 & 100 & 100 & 3 \\
    CAT petal to SPO petal & 0.333 & 0.333 & 0.333 & 100 & 100 & 200 \\
    CAT windows to CAT petal & 0.333 & 0.333 & 0.067 & 100 & 60 & 100 \\
    individual CAT to window & 0.333 & 0.333 & 0.067 & 100 & 60 & 100 \\
    Camera to front assembly & 1.667 & 0.667 & 0.333 & 60 & 60 & 60 \\
    \end{tabular}
\end{table}

\begin{figure}
    \centering
    \includegraphics[width=0.75\textwidth]{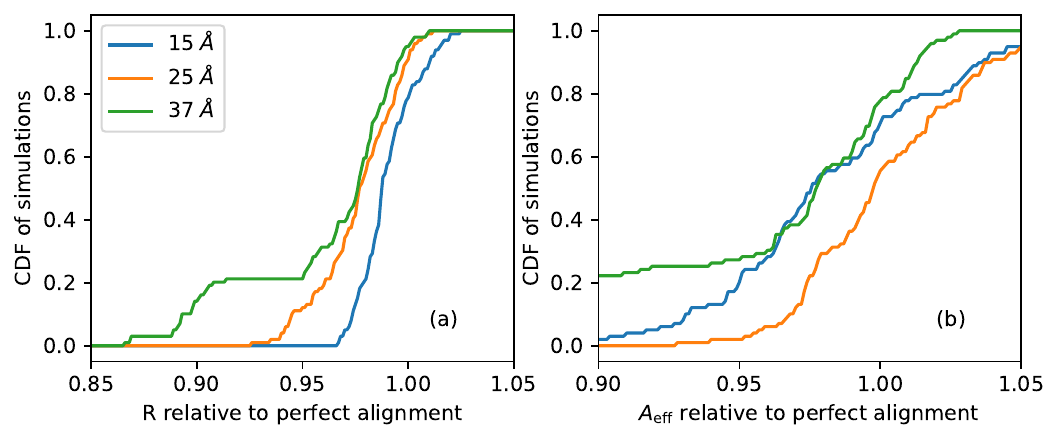}
    \caption {\label{fig:alignment_budget}
    Cumulative distribution function for $R$ (panel a) and effective area (panel b) relative to a perfectly aligned Arcus for a set of simulations where all elements are simultaneously misaligned according to the misalignment budget in Table~\ref{tab:tolerances}.}
\end{figure}

The alignment budget assumes that all alignment tolerances contribute independently. The next set of simulations is designed to check this assumption. Some misalignments might cancel out in practice, others might have a multiplicative effect. Full ray-tracing is the best way to check that and to predict final instrument performance. We run a set of 100 ray-traces, and draw a new set of mislaignments from Table~\ref{tab:tolerances} for each of them. Figure~\ref{fig:alignment_budget} shows the cumulative distribution of $R$ and effective area for those simulations relative to a perfectly aligned Arcus.

Differences are visible between the different wavelengths. At 37~\AA{} two dispersed orders (order -3 and -4) contribute, but order -3 is close to the edge of a CCD and order -4 is close to a chip gap. If either one is lost, $R$ or $A_\mathrm{eff}$ at that wavelength will suffer, but at the same time a neighboring wavelength will improve because it is no longer inside a chip gap. In contrast, the signal at 25~\AA{} is dominated by order -6, comfortably in the middle of a CCD. Thus, the two plots above should not be interpreted as ``longer wavelength will suffer more''; crucial orders are close to a chip gap in different spots over the Arcus bandpass. Instead, the plots should be read as showing the range of effects that the baseline misalignment can have on $R$ and $A_\mathrm{eff}$, depending on what exactly the random numbers are that are drawn.

Apart from the effects of shifting chip gaps, Figure~\ref{fig:alignment_budget} shows that the alignment tolerances listed in Table~\ref{tab:tolerances} give us 95\% of the nominal $R$ and effective area in more than 90\% of the realizations.

\section{X-RAY INSTRUMENT PARTICLE BACKGROUND}
\label{sect:nxb}

Galactic cosmic ray (GCR) particles interact with the Arcus X-ray instrument camera structure, generating secondary X-rays and electrons that produce signals in the detectors that are indistinguishable from focused celestial X-rays. To first order, this background is uniform across the focal plane due to the isotropic GCR flux, and therefore it has a very different spatial-spectral pattern compared to the diffracted spectra produced by the optics and gratings.

To simulate the effect of this non-X-ray background (NXB) on per-order spectral analysis of Arcus X-ray observations, we produced a model spectrum using in-flight data from CCD detectors on Chandra (ACIS-S3) and Suzaku (XIS1). These heritage detectors share many features with the Arcus X-ray CCDs: identical pixel size (24 $\mu$m), backside illumination with similar depletion depth (45 $\mu$m for ACIS-S3 and XIS1 vs.\ 60 $\mu$m for Arcus), and similar operational and on-board processing parameters. The Chandra ACIS-S3 NXB spectrum was derived from ACIS stowed data taken in 2005--2009, during a time of intermediate Solar activity. The Suzaku XIS1 spectrum was derived from observations of the dark (night) Earth limb taken over the course of the full 2005--2015 mission. Both datasets were filtered in standards ways, with no Cut-Off Rigidity (COR) filtering applied for Suzaku\cite{Tawaetal2008}.

As expected from the detector similarities, the spectra in Figure~\ref{fig:nxb} (left panel) show very similar features: a general power-law slope; instrumental fluorescence lines of O K (0.5~keV), Al K (1.5~keV), Si K (1.7~keV), and Au M (2.3~keV); and upturns at low and high energies. The Chandra NXB is about a factor of four higher than Suzaku when scaled to physical detector area, expected due to the different particle environments in high- and low-Earth orbits. Arcus should have a very similar overall unrejected background event rate to Chandra ACIS, as it will also be outside the Earth's protective magnetic field. There are some important changes we implemented to model the Arcus NXB spectrum:

\begin{figure} [t]
    \begin{center}
    \begin{tabular}{c} 
    \includegraphics[width=.5\textwidth]{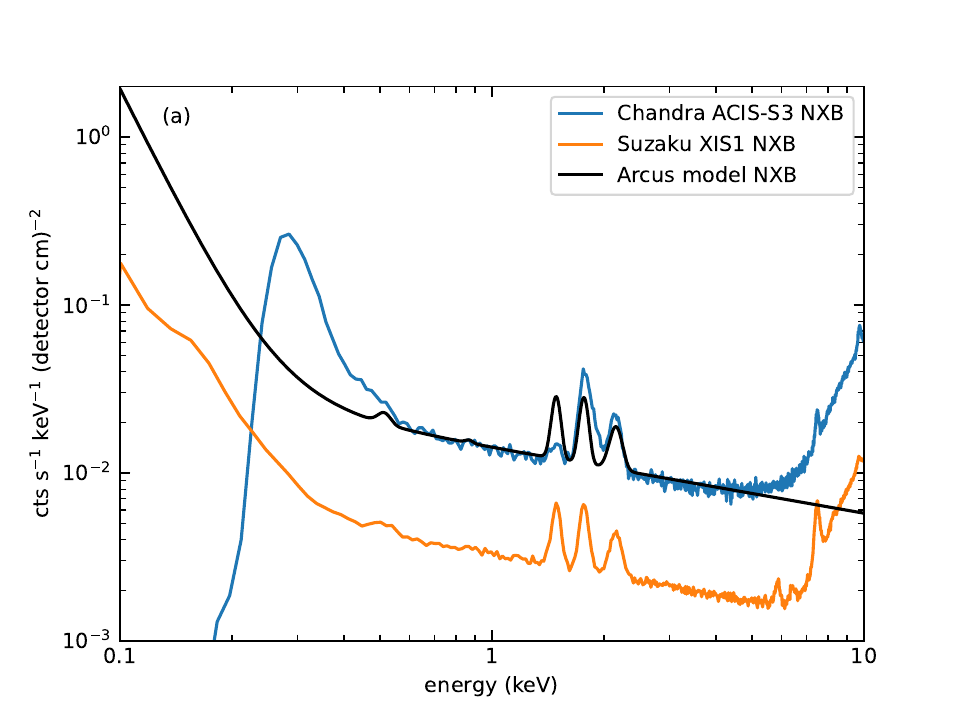}
    \includegraphics[width=.5\textwidth]{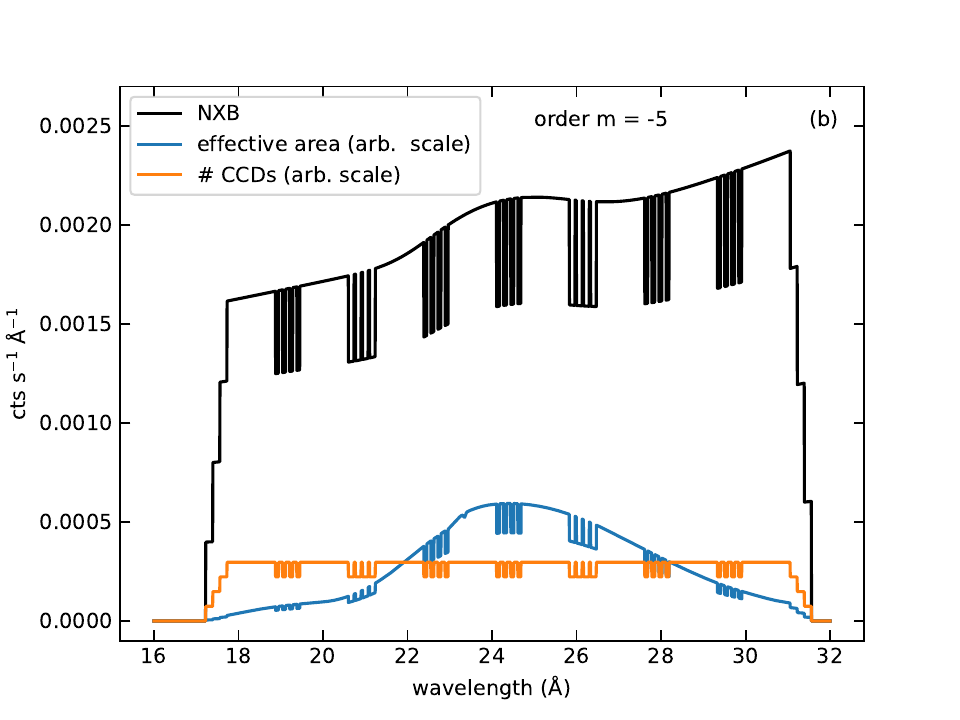}
    \end{tabular}
    \end{center}
    \caption {\label{fig:nxb}
    (panel a) Observed non-X-ray background (NXB) spectra for the Chandra ACIS-S3 and Suzaku XIS1 CCD detectors, shown with the adopted Arcus model spectrum. XIS1 has a lower background due to Suzaku's low-Earth orbit, compared to the high-Earth orbits of Chandra and Arcus. Other features are discussed in the text. (panel b) Example simulated per-order NXB spectrum, shown along with the number of CCDs and effective area (both with arbitrary scaling) as a function of wavelength. The broad feature peaking near 24 \AA\ in the NXB spectrum is due to the O K fluorescence feature seen near 0.5 keV in panel (a). This demonstrates the subtle effects that the NXB and detector energy response can impose on extracted per-order spectra.
    }
\end{figure}

\begin{enumerate}
    \item The upturn at high energy is removed. This upturn is due to cosmic ray minimum-ionizing particles (MIPs) that hit the detector at normal incidence and traverse the full detector thickness, confined to a single pixel. The distribution of energy each MIP deposits is governed by the detector thickness, but it is quite broad, extending to low energies. Since the Arcus CCD is somewhat thicker (60 $\mu$m) than ACIS-S3 (45 $\mu$m), this feature will move to higher energies above the Arcus band.
    \item The upturn at low energies from Suzaku is used instead of Chandra. It is likely that the gain scale at low energies for ACIS-S3 is non-linear due to high charge-transfer inefficiency (CTI), and we suspect events here suffer incorrect energy assignment. The Arcus CCD performance at low energies will be much closer to XIS1.
    \item The instrumental fluorescence lines will likely differ in amplitude for Arcus, but we have kept the Suzaku line strengths to be aware of their presence.
\end{enumerate}

In practice, we have simply fit an empirical model to the Suzaku spectrum, ignoring the high-energy upturn, and scaled that to match the ACIS-S3 rate at 1 keV.

To create per-order NXB spectra, the CCD NXB model was first converted to cts s$^{-1}$ keV$^{-1}$ \AA$^{-1}$ by multiplying by the spatial cross-dispersion width of the extraction window (50 pixels = 1.2 mm) and by the dispersion relation (in \AA\ mm$^{-1}$) inferred from the ray-tracing described in Section \ref{sect:raytracing}.  At each response channel wavelength, we integrate the CCD NXB model over a window 70 eV wide, centered on that energy. This results in a spectral count rate in units of cts/s in bins equivalent to those in the per-order response. Finally, we multiply the rate by the number of CCDs contributing to the sensitivity at each wavelength. Example NXB spectra are shown in the right panel of Figure \ref{fig:nxb}; these can be used with per-order Arcus response files to simulate the effects of the NXB on the analysis of faint science targets.

\section{SUMMARY}
\label{sect:summary}

Arcus is a Probe-class mission for FUV and soft X-ray spectroscopy with unprecedented resolving power and effective area. We discuss potential sources of misalignment between the UV and X-ray spectrographs and procedures to calibrate and mitigate any differences in line-of-sight. Using ray-tracing, we predict the Arcus performance and the effect of misalignments, which sets the limits for alignment tolerances with in the XRS. We also discuss the non-X-ray background and its effect on the Arcus performance.

\acknowledgments 
Support
for this work was provided in part through NASA grant NNX17AG43G and
Smithsonian Astrophysical Observatory (SAO) contract SV3-73016 to MIT
for support of the {\em Chandra} X-Ray Center (CXC), which is operated
by SAO for and on behalf of NASA under contract NAS8-03060.  The
simulations make use of Astropy, a community-developed core Python
package for Astronomy\cite{astropy:2022}, numpy\cite{numpy}, and IPython\cite{IPython}.

\bibliography{report} 
\bibliographystyle{spiebib} 

\end{document}